\documentclass[runningheads]{llncs}
\usepackage{graphicx}
\usepackage{todonotes}
\usepackage{amsfonts}
\usepackage{multirow}
\usepackage{booktabs}
\usepackage{subfiles}
\usepackage{anyfontsize}

\usepackage[caption=false]{subfig}

\usepackage[symbol]{footmisc}

\let\oldmaketitle\maketitle
\renewcommand{\maketitle}{\oldmaketitle\setcounter{footnote}{1}}
\begin{document}
	\title{A Convolutional Approach to Vertebrae Detection and Labelling in Whole Spine MRI}
	\titlerunning{Convolutional Vertebrae Detection and Labelling}
	%
    \author{Rhydian Windsor\inst{1} \and Amir Jamaludin \inst{1} \and Timor Kadir \inst{2} \and Andrew Zisserman \inst{1}}
	%
	\institute{Visual Geometry Group, Department of Engineering Science, University of Oxford \and  Plexalis Ltd}
	\authorrunning{R. Windsor et al.}
	%
	%
	\maketitle              
	\begin{abstract}
       We propose a novel convolutional method
        for  the detection and identification of vertebrae in whole spine
        MRIs.
        This involves using a learnt vector field to group detected vertebrae corners together into individual vertebral bodies and convolutional image-to-image translation followed by beam search to label vertebral levels in a self-consistent manner.
        The method
        can be applied without modification to lumbar,
        cervical and thoracic-only scans across a range of different MR
        sequences.
		The resulting system achieves 98.1\% detection rate and 96.5\% identification rate on a challenging clinical dataset of whole spine scans and matches or exceeds the performance of previous systems of detecting and labelling vertebrae in lumbar-only scans. Finally, we demonstrate the clinical applicability of this method, using it for automated scoliosis detection in both lumbar and whole spine MR scans.
	\end{abstract}
	\section{Introduction}
	
The objective of this paper is automated vertebrae  detection and identification of vertebral levels.
This is an important task for several reasons. Firstly, automated diagnosis of many spinal diseases such as disc degeneration~\cite{Jamaludin17b,Jamaludin17}
or spinal stenosis~\cite{Lu18spine} relies on accurate localisation of vertebral structures or, in the case of pathological scoliosis, 
lordosis and kyphosis~\cite{Jamaludin19a}, analysing the geometry of the spinal column. 
Secondly, vertebral bodies can be used to infer other spinal structures of interest such as the spinal cord or ribs. 
Finally, vertebrae can act as points to allow registration between different scans.

There are several issues that make this task
challenging.  One of the most obvious is that vertebrae are highly
repetitive and hence distinguishing between different levels can be
hard. Labelling by simply counting down from the C2 vertebra is problematic as it assumes that all vertebrae have been
detected, C2 is visible, and that every patient has the same number of
vertebrae which is not always true. Furthermore, for clinical use, labelling must be
robust to: variations in spinal anatomy (such as collapsed vertebrae, hemivertebrae and fused vertebra); vertebrae numbers
--  around 11.3\% of the
population have one more or one less mobile vertebra~\cite{Tins16}); 
different imaging parameters including MR weighting (e.g.\ T1, T2,
STIR, TIRM and FLAIR); fields of view (e.g. lumbar,
whole spine scans); scan resolution and also number/thickness of
slices in the scan.  

This paper proposes a new approach to this challenge, in particular in
the case of 3-D sagittal {\em whole spine} clinical MRIs which are
important for diagnosing several diseases such as
ankylosing spondylitis and multiple myeloma.  
We make the following contributions to the tasks of {\em vertebral body detection} and {\em vertebral level labelling} in clinical MRIs:
We propose a new convolutional method of detection based on localising the corners and centroids of vertebrae and then grouping them together (Section~\ref{sec:detection_methods});
We reformulate the labelling task as a convolution enhancement followed by a language modelling inspired sequential correction, removing the need for a recurrent network and showing robustness to variations in vertebra numbers (Section~\ref{sec:labelling_methods}); 
We show that the resulting system is robust to a variety of fields of view and pathologies by evaluating on a large clinical dataset of MR lumbar scans, 
and also, for the first time, on whole spine scans. We achieve state-of-the-art performance at vertebra identification in both (Section~\ref{sec:detection_and_labelling_results}); 
We demonstrate a clinical application of this system by using it to
automatically detect cases of scoliosis (Section~\ref{sec:scoliosis}).

 \noindent\textbf{Related work:} There have been several approaches to automated detection and labelling of vertebrae in MRIs although most focus on fixed fields of view, e.g. lumbar or cervical scans only \cite{forsberg_detection_2017,Jamaludin17b,Lootus14,Lu18spine}. Previous vertebrae labelling methods tend to rely on either heuristic-based graphical models \cite{forsberg_detection_2017,Lootus13} or assuming the bottom vertebra is S1 and counting up \cite{Lu18spine}. Zhao et al. \cite{zhao_automatic_2019} perform labelling in MRIs with arbitrary fields of view, but only in the lower spine (from S1-T12 to L4-T10). Windsor and Jamaludin~\cite{Windsor20} report a method of detecting, though not labelling,  vertebrae in full spine scans iteratively but also require the location of the S1 vertebra for initialization. Cai et al. \cite{cai_multi-modality_2015} also perform detection and labelling by a hierarchical deformation model and even report success in a single full spine scan but do not evaluate the performance quantitatively on a whole spine dataset. Furthermore such models are slow to apply and make strong assumptions on spinal geometry such as a fixed number of vertebrae and lack of major pathology (fused/collapsed vertebrae). Greater progress has been made in whole-body CT images where the task is more straightforward; CT imaging protocols are highly standardised due to image intensities being consistent, representing X-Ray absorptions of voxel locations. Also, CT scans tend to have more 3-D information, with higher resolution in the sagittal plane than that typical of clinical MR. Approaches using graphical models and recurrent neural networks for labelling have been reported to achieve 70-89\% identification rate in mixed field of view CT scans \cite{glocker_automatic_2012,glocker_vertebrae_2013,yang_automatic_2017}.
\section{Detecting Vertebrae}
\label{sec:detection_methods}

\begin{figure}[!t]
		\centering
		\includegraphics[width=\textwidth]{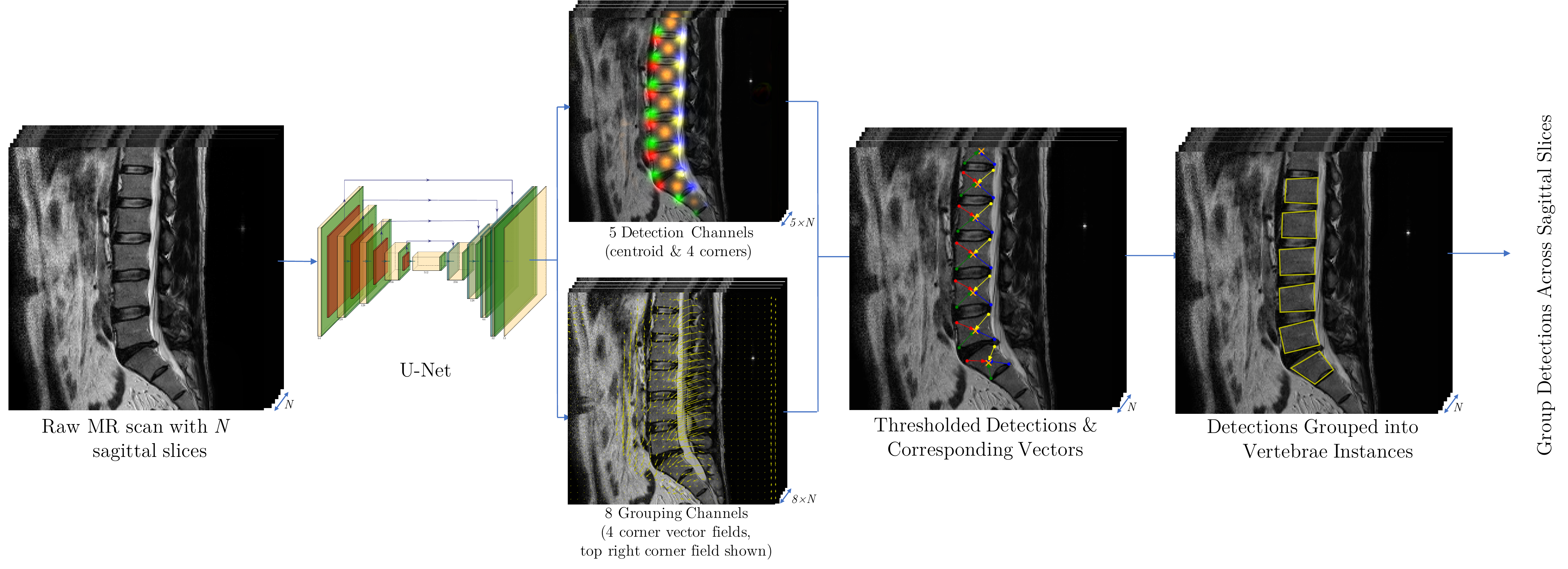}
		\caption{The pipeline used to detect vertebrae in whole spine scans. The output detection channels are thresholded and each resulting connected volume becomes a corner or a centroid.  Each corner has corresponding vector in the grouping channels at the point-of-detection and `points' with a magnitude and direction according to that vector. Each centroid is assigned the 4 corners that point closest to it.}
		\label{fig:detection_pipeline}
	\end{figure}

Detection of the vertebrae proceeds in three stages, as illustrated in
Figure~\ref{fig:detection_pipeline}. 
First, corners and centroids of each vertebra are predicted in each sagittal slice. Second, 
each detected corner is assigned to a centroid by predicting a vector
field for each corner type (e.g.\ top left,  bottom right) which points to the corresponding
centroid. The detected centroid  and the four corners which point
to it  define the bounding quadrilateral for that vertebra in
that slice. Third, these quadrilaterals are grouped across slices to define detected volumes for  each vertebra.
	
Inference is performed here by a U-Net architecture (shown in detail in the appendix) which ingests an MRI (one slice) and outputs 13 channels: 5 channels are the \textit{landmark detection} channels and each is used to detect the centroid and 4 corners respectively of all vertebral bodies appearing in that image; the remaining 8 channels are the \textit{landmark grouping} channels corresponding to the $x$ and $y$ coordinates of the 4 vector fields used in grouping.\newline\indent In more detail, the locations of corners and centroids are identified as modes of the heatmaps in the \textit{landmark detection} channels. For each corner detected, there exists a corresponding grouping vector field from the \textit{landmark grouping} channels; 4 corners equals 4 vectors fields. A group is then formed from the 4 closest corners pointing to a single centroid; a group here forms a quadrilateral. This is done for each detected centroid. If two or more centroids are assigned the same corner, the centroid to which the corner's vector
	points closest remains, and the other centroid is discarded to stop double detection of a single vertebra. If there is no detected corner	within a fixed range of a centroid, it is also discarded, making the system robust to spurious centroid detections. This is performed
	in each sagittal slice. Finally, the vertebral bodies detected are grouped across slices by measuring the IoU between quadrilaterals and assigning 
	them to the same vertebra if they have an overlap of greater than 0.5.

\noindent\textbf{Discussion.} Our approach of detecting vertebrae as a series of points and then grouping them differs from other methods which have used region proposal networks \cite{zhao_automatic_2019}
	pixel/voxel-wise segmentation \cite{zukic_robust_2014}, deformable-part models \cite{cai_multi-modality_2015} or simply detected vertebrae centroids \cite{forsberg_detection_2017,yang_automatic_2017}.
	However, this approach is receiving increasing attention in the computer vision literature, with Zhou et al. \cite{Zhou19} showing state-of-the-art results by detecting objects as a series of keypoints. The advantage of this method is its high speed (1 second inference on GPU-enabled hardware) combined with more accurate bounding regions than using standard bounding boxes.

\noindent\textbf{Training:} To train the network, we use separate loss functions for the detection and grouping channels. The ground truth annotations are coordinates of the corners of the vertebral bodies (centroids are computed from these). Target detection channel outputs are constructed by overlaying a Gaussian kernel on
each annotated ground truth landmark in the detection channels with
variance proportional to the square root of the area of the
detection. These channels are trained using the weighted L1 loss; $\mathcal{L}_{detect}(Y, \hat{Y}) = \sum_{k=1}^5 \sum_{i,j} \alpha_{ijk}|y_{ijk} - \hat{y}_{ijk} |$
	where $Y$ is the response map output by the network and $\hat{Y}$ is the target response map. $y_{ijk}$ and $\hat{y}_{ijk}$ are the value of response and ground truth map respectively at image coordinate $(i,j)$ in detection channel $k$ and $a_{ijk}$ is a weighing factor given by $ \alpha_{ijk} = \frac{N_k}{N_k + P_k}  \mbox{if } \hat{y}_{ijk} \geq T$ or $
	\frac{P_k}{N_k + P_k} \mbox{if } \hat{y}_{ijk} < T$,
	where $P_k$ and $N_k$ are the number of pixels respectively above and below threshold $T$ in channel $k$. This weighing factor balances the loss from false positive responses in the heatmap and false negatives, speeding up training. In the experiments in this paper, $T=0.01$ was used. \textit{Landmark grouping} channels are trained by using the L2 loss;
	$
    	\mathcal{L}_{group} = \sum_{l=1}^4\sum_{b}\sum_{(i, j) \in \mathcal{N}_{bl}} ||\mathbf{v}^{l}_{ij} - \mathbf{r}^k_{ij}||_2^2.
    $
	Here $l$ indexes each corner type/vector field (e.g.\ top left, bottom right), $b$ indexes labelled ground truth vertebral body, $\mathcal{N}_{bl}$ is a neighbourhood of pixels surrounding the $l^{th}$ corner of vertebral body $b$. $\mathbf{v}^l_{ij}$ is the value of the vector field corresponding to corner $l$ at the pixel location $(i,j)$ and $\mathbf{r}^b_{ij}$ is the displacement from the centroid of vertebral body $b$ to $(i, j)$. As suggested by \cite{ronneberger_u-net_2015}, heavy augmentation is used during training. Scans are padded, rotated, zoomed and flipped in the coronal plane. Non-square scans are split into squares overlapping by 40\% and resized to $224\times224$ to ensure constant sized input to the network.
	\section{Labelling Vertebrae}
	\label{sec:labelling_methods}
\begin{figure}[!t]
		\centering
		\includegraphics[width=\textwidth]{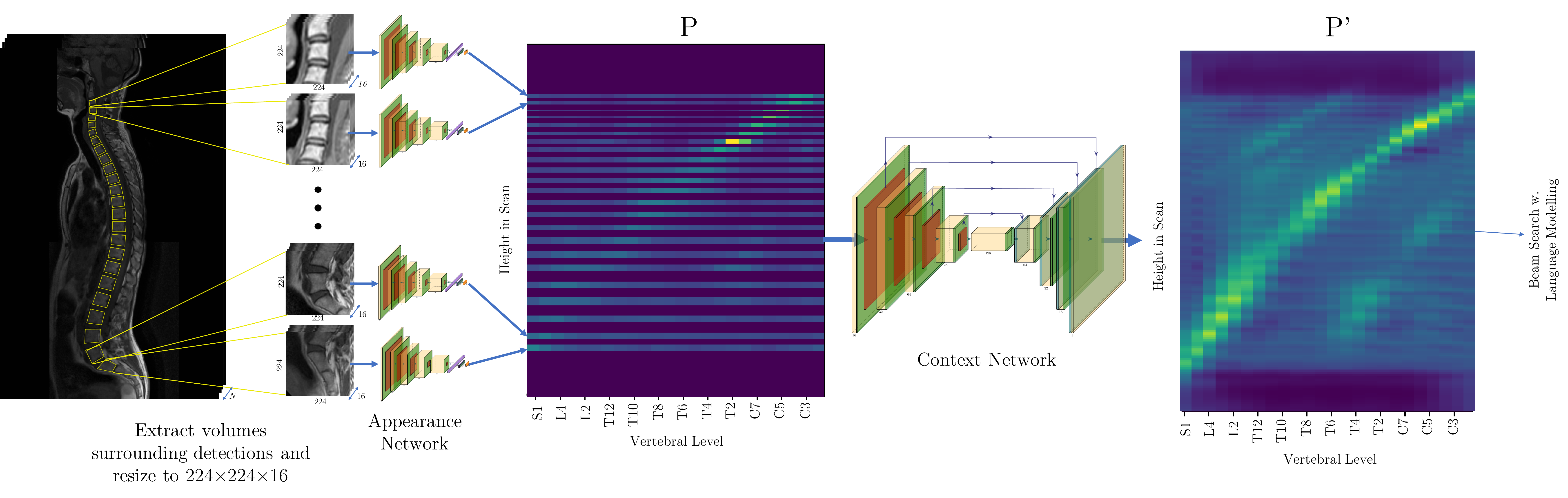}
		\caption{Overview the labelling pipeline. Volumes surrounding each of the vertebrae detected are extracted and input to an appearance network. This is then used to construct an input image to the context network which gives the probability of a given label at a given height in the image. The resulting probability map is shown in the far right of the image, followed by a beam search to generate the final level sequence. The architectures of both networks are shown in the appendix.}
		\label{fig:labelling_pipeline}
	\end{figure}

Once the vertebrae are detected, the next task is to label each vertebrae with its  level (e.g.\ S1, L5, L4 etc.). 
There are two types of information to consider when labelling a vertebra; the \textit{appearance} of the vertebrae -- its intensity pattern, shape and size -- and its \textit{context} -- its position in relation to
other vertebrae that have been detected in the scan. As such, we train two networks; an \textit{appearance network} to infer the level 
	of vertebra from its appearance alone, and a \textit{context network} which takes as input the predictions
	of the appearance network along with the spatial configuration of the detections in the scan to improve the predictions. The final stage is to search for a consistent labelling of the vertebra using a sequential `language model' that builds
in ordering constraints.
The labelling pipeline is outlined in Figure~\ref{fig:labelling_pipeline}. It should be noted that both networks are fully convolutional. 
	This differs from the approaches outlined in~\cite{yang_automatic_2017} and~\cite{Lootus14} which use recurrent neural networks and graphical models respectively. 
	
The labelling pipeline proceeds as follows: A volume around each detected vertebra is extracted and given as input to the appearance model which predicts a softmax probability vector over the labels.
Input volumes are created by first fitting a bounding box around the
detection, then expanding the box by 100\% to include nearby
anatomical features and resampling its size to $224\times224\times16$,
where 16 is the number of sagittal slices. 
For the appearance network, we use a simple 3D CNN (details of this and the context network
are given the appendix).
Then, the predictions of the appearance networks are used to make a probability-height map where the probability vector at the height of each vertebra is equal to the output vector of the appearance network for that vertebra.
In detail: the probability-height map, $P\in \mathbb{R}^{H\times N}$ where $H$ is the height of the image and $N=24$ is the number of vertebrae level classes (C2 to S1). If vertebra detection $v$ spans from heights $h_{v_1}$ to $h_{v_2}$ in the scan then $P_{h'n} = p^{a}_{vn} \forall \; h_{v_1} \leq h' \leq h_{v_2} $ and 0 otherwise where $p^a_{vn}$ is the probability $v$ has level $n$, given by the appearance network.
Next, the context network refines this probability-height map, using detections around each vertebra to update its predicted class.
Finally, the output map is decoded into a logical sequence of vertebra using a language-model  inspired beam search which imposes global constraints such as no repetitions. This is described in detail in Section \ref{sec:language_modelling}.

\noindent\textbf{Training: }The appearance network is trained using a cross-entropy loss function. For re-calibration \cite{guo_calibration_2017}, a softmax operation with temperature $T=10$ is applied to the logits layer. $P$ is given to the context net which produces $P'$, a refined version of the same map as shown on the far right of Figure \ref{fig:labelling_pipeline}. The cross-entropy between the output probability map at the centroid height of each vertebra and its ground truth label is used as the loss function. When training the context network, augmentation is applied at training time by randomly removing between 0 to 4 of the highest and lowest detections from the input image and further removing each remaining vertebra's probability map from the input image with probability 0.2. This ensures the labelling pipeline is robust to missed detections and not reliant on distinct looking vertebrae at the top and bottom of the  spine.
	\subsection{Enforcing Monotonicity and Dealing with Numerical Variations}
	\label{sec:language_modelling}
To predict the labels for each vertebrae detected, $P'$  shown in 
Figure \ref{fig:labelling_pipeline} must be decoded into a sequence of level predictions. 
A naive method to do this would be to take the argmax of the probability map at each vertebra centroid height and use this as a label. However, this allows implausible sequences such as repeated vertebral levels to be predicted. If the detection $v$ at height $h_v$ has label index $l$, then we know $v'$ at $h_{v'}>h_v$ detection should have label $l'>l$ with the greatest probability of being $l+1$(e.g. L4 should be followed by L3). 
The challenge of imposing such constraints is analogous to that faced
in automatic speech recognition where CTC training is followed by language modelling to predict a valid character or
word sequence~\cite{scheidl_word_2018,graves_connectionist_2006}. We take inspiration from this
and use a beam search to obtain a valid labelling:
beginning at the highest detection and
searching down, sequences are generated and scored by selecting the highest probability labels for
each vertebra.
The $k$ most likely sequences of levels are stored in
memory at each step, where $k$ is the beam width. Sequences with
repetitions of levels are given probability 0 and those with skipped
levels are penalised by multiplying the sequence probability by
a penalty score. The method can also incorporate numerical variations, with $\pm1$ lumbar vertebrae given a small sequence
probability penalty reflecting the low incidence of this.

\section{Detection and Labelling Results}
	\label{sec:detection_and_labelling_results}
We evaluate the system at the task of vertebrae detection and labelling in whole spine and lumbar scans. 
Following ~\cite{Lootus13,yang_automatic_2017}, we define correct detections to be when the ground truth vertebra centroid is contained entirely within a single bounding quadrilateral. For detection, we report precision, recall,  and the localisation error (LE), defined to be the mean distance of ground truth centroids from the closest detection quadrilateral centroid. For labelling, we report identification rate (IDR), the fraction of vertebrae detected and labelled correctly.
	\subsection{Datasets}
	Three datasets are used in this work: OWS, Genodisc and Zuki\'c. 
	OWS is a dataset of 710 sagittal whole spine scans across 196 patients from the Picture Archiving and Communication System (PACS) of an orthopaedic centre. The dataset exhibits a wide range of pathologies such as hemivertebra, fused vertebrae, numerical variations of vertebrae and scoliosis.
	Scans are taken from different scanners with a range of MR parameters (T1, T2, FLAIR, TIRM and STIR). 
	The dataset is split into training, validation and testing sets with an 60/20/20\% split at the patient level. Corners of vertebrae from S1 to C2 are annotated and vertebrae levels were marked by a radiologist in one scan for each patient, with S1 being the first vertebra attached to the pelvis. 
	In the case of 25 vertebrae between S1 and C2 instead of the normal 24, an extra lumbar vertebrae is labelled (L6).
	Networks were trained on the OWS training set, using an Adam optimizer with $\beta_1=0.9$, $\beta_2=0.999$ and a learning rate of 0.001.
	The Genodisc and Zuki\'c datasets are used only for testing. Genodisc's test set has 421 clinical lumbar MRIs used by Lootus et al.~\cite{Lootus14}. Zuki\'c~\cite{zukic_robust_2014} is a small dataset of 17 mostly lumbar sagittal MRIs available on the online SpineWeb platform.
	\subsection{Results}
	The results of detection and labelling on all datasets are shown in Table~\ref{tab:detection_results} with comparisons to other methods reported on the same datasets where available. We also compare our convolutional labelling pipeline results to a baseline recurrent approach for vertebra labelling, training a bidirectional LSTM on the appearance features extracted from each detected volume. Example predicted detection and labelling sequences across a range of pathologies are given in Figure \ref{fig:pathologies}. The LSTM baseline used is detailed in the appendix.
	
	\begin{table}[!t]
		\centering
		{\fontsize{7.55}{9}\selectfont 
		\begin{tabular}{c|c|c|c|c|c|c|c|c}
		
			\toprule
			Dataset & No. Scans & No. Vert & Method & Prec. (\%) & Rec. (\%) & IDR(\%) & IDR$\pm$ 1(\%) & LE (mm)  \\
			\toprule
			
			 \multirow{3}{*}{\shortstack{OWS \\ (Whole \\Spine)}}& & & Windsor\textsuperscript{\textdagger}~\cite{Windsor20}& \textbf{99.4} & \textbf{99.4}  & - & - & \textbf{1.0 $\pm$ 0.9} \\
			 & 37 &  888 & Label Baseline & - & - & 86.9 & 93.4 & - \\
			 & & & Ours & 99.0 & 98.1 & \textbf{96.5} & \textbf{97.3} & 2.4 $\pm$ 1.3 \\
			\hline
			\multirow{3}{*}{\shortstack{Genodisc \\ (Lumbar)}}& & &  Lootus~\cite{Lootus13}& - & - & 86.9 & - & 3.5 $\pm$ 3.3\\
			 & 421 & 2947 & Label Baseline & - & - & 90.1  & 97.4 &  -\\
			 & & & Ours & \textbf{99.7}  & \textbf{99.7}  & \textbf{98.4} & \textbf{99.7} & \textbf{1.6 $\pm$ 1.1}\\
			\hline
			\multirow{3}{*}{\shortstack{Zuki\'c \\ (Lumbar)}}& & & Zuki\'c~\cite{zukic_robust_2014}& 98.7 & 92.9 & - & - & \textbf{1.6 $\pm$ 0.8}\\
			& 17 & 154 & Label Baseline & - & - & 87.0 & 94.3 & -\\
			  & & & Ours & \textbf{99.3}  & \textbf{98.7}  & \textbf{90.9} & \textbf{98.7} & 2.0 $\pm$ 1.5\\
			\hline
		\end{tabular}
	    }
\caption{Performance of the pipeline on the three datasets. Our approach is compared with other methods using the same datasets and also a LSTM labelling baseline. Results are reported on a per-vertebra level. 
Higher is better for detection precision (Prec.), detection recall (Rec.),  and level identification rate (IDR).  Lower is better
for localisation error (LE). We also report the percentage of vertebrae within one level of their ground truth value (IDR$\pm1$). Lootus \cite{Lootus13} is tested on a subset of 291 scans from the Genodisc dataset.{\bf Note, Windsor\textsuperscript{\textdagger}~\cite{Windsor20} requires manual initialization by providing the location of the S1 vertebra, so is not directly comparable}} \label{tab:detection_results}
	\end{table}
	
	\begin{figure}
	    \centering
	    \includegraphics[width=0.9\textwidth]{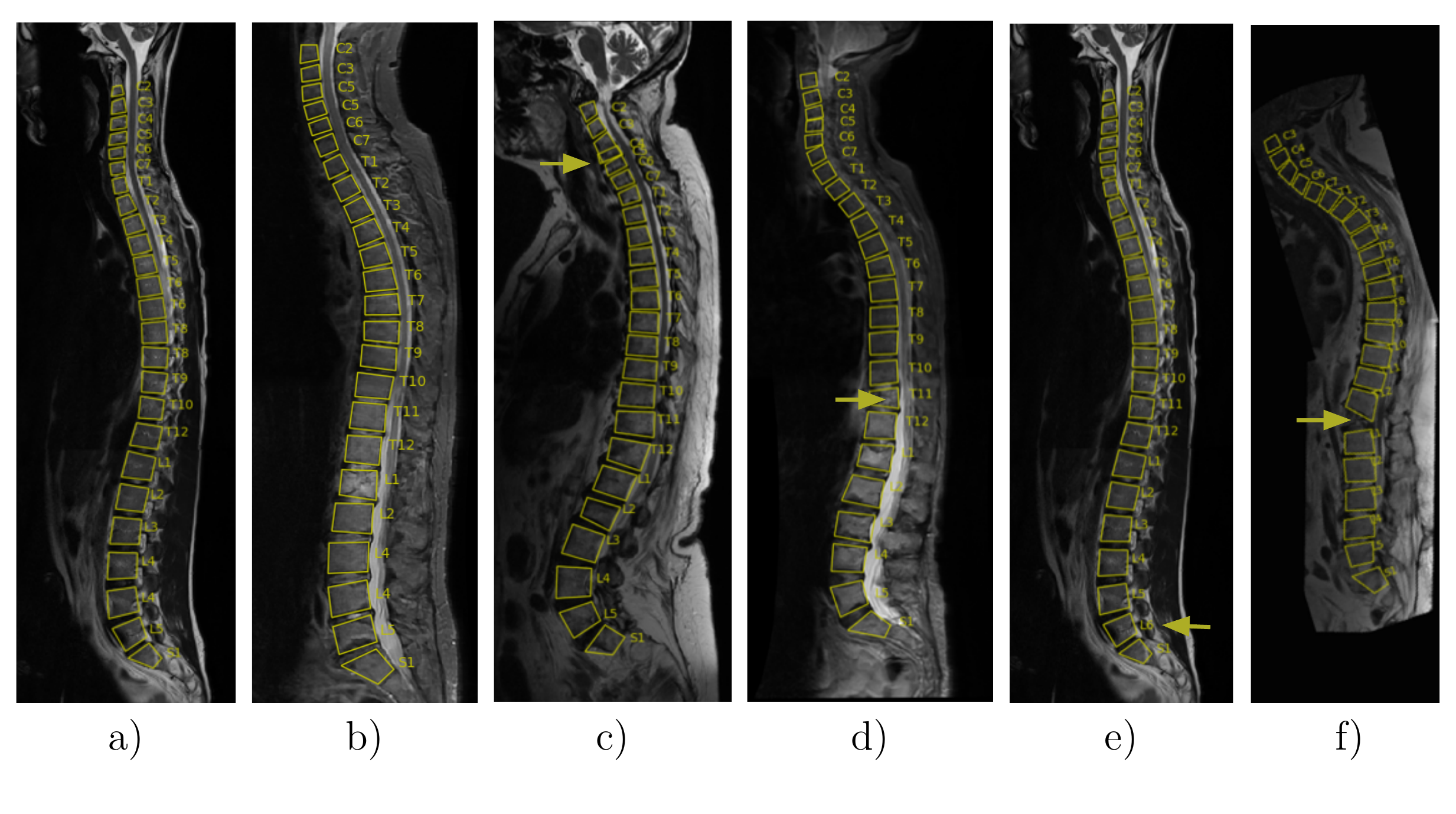}
	    \caption{Example detection and labelling of vertebrae for a range of whole spine scans. a) and b) are examples of typical spines. The other images shows examples of the system dealing with pathologies; c) shows an example of a spine with fused C3 and C4 vertebrae; d) is an example of a spine with a collapsed vertebra; e) shows a spine with an extra vertebra between S1 and C2 and f) shows an extremely kyphotic spine with a hemivertebra at the bottom of the thoracic spine.}
	    \label{fig:pathologies}
	\end{figure}
\noindent\textbf{Discussion:} 
For whole spine detection on the OWS dataset, the proposed method achieves a high precision and recall of
99.0\% and  98.1\% respectively. It achieves a level identification rate of 96.5\%, significantly exceeding 
the LSTM baseline.
The few errors the system made
for labelling are generally due to S2 being detected as S1, meaning
all labels are out by one. In practice, this is a mistake
radiologists often make as it can be difficult to tell which is the
first sacral vertebra without looking at axial scans to see which bone
is joint to the pelvis. This also explains why \cite{Windsor20}
achieves slightly higher precision and recall at detection than ours;
it is a semi-automated algorithm given the location of S1 at
initialization and thus bypasses this difficult problem of S1
recognition. 
On the Genodisc lumbar spine dataset, our method again significantly outperforms the baseline,
and it is also outperforms the prior method of Lootus et al.~\cite{Lootus13} (98.4\% compared to 
86.9\%).
Importantly, OWS and Genodisc are scans of patients with a
wide range of pathologies imaged using typical clinical MR protocols;
hence strong performance here gives evidence of the clinical
usefulness of this approach. We show example results for pathological
spines in the appendix. In the appendix we also report results by other
groups for vertebrae detection and labelling in MRIs. However, these
are for different datasets to which we could not get access, with
different scanning protocols, fields of view (FoV) and patient sets,
and thus cannot be compared to directly. 
\section{Automated Scoliosis Detection}
\label{sec:scoliosis}
\begin{figure}[!t]
	\centering
	    \includegraphics[width=0.9\textwidth]{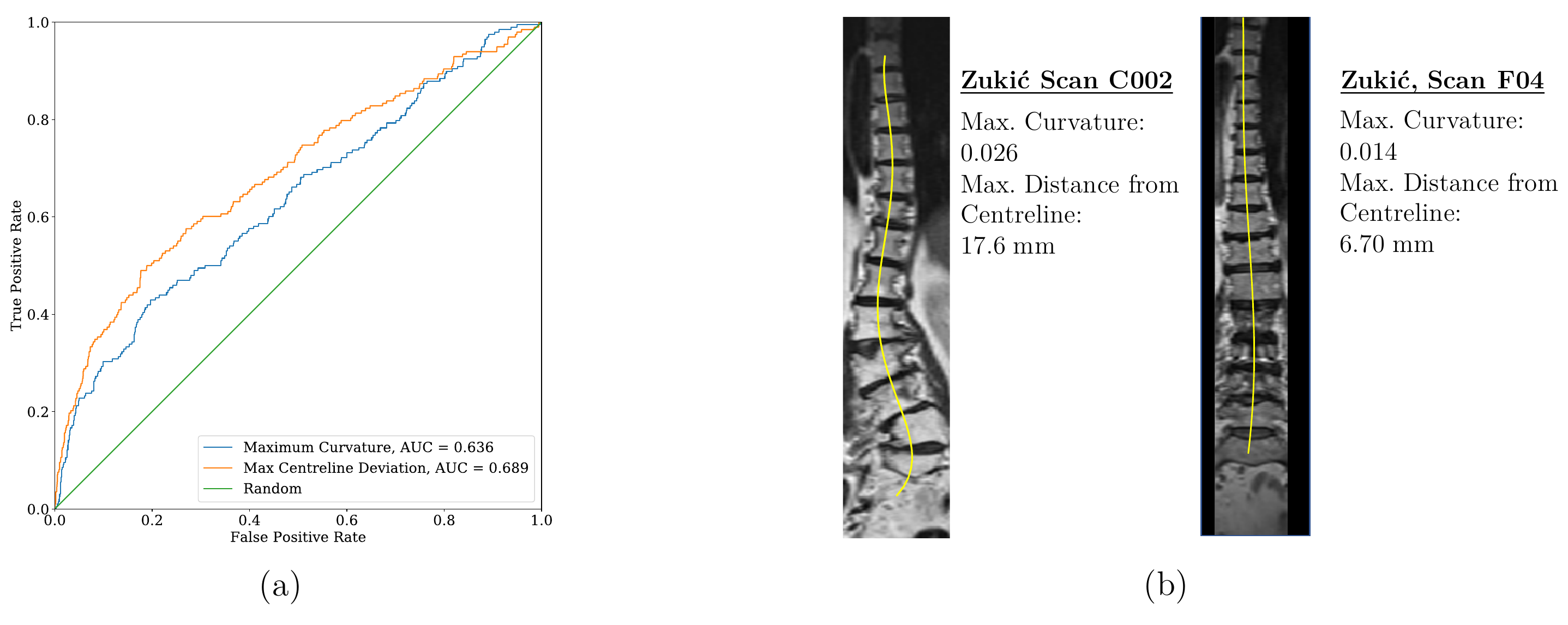}	
		\caption{Results of  scoliosis classification in sagittal scans using the proposed vertebrae detection system: (a) ROC  curve for simple scoliosis classifiers based on statistics of polynomial curves fitted through vertebrae centroids in Genodisc scans; (b) A qualitative comparison of curves fit through detected vertebrae in full spine scans. Curves are overlaid on coronal slices synthesised from the sagittal slices. }
		\label{fig:scoliosis}
	\end{figure}

Finally, as an illustration of a potential application of the proposed approach, we explore the ability of the system to classify cases of scoliosis from sagittal MR scans. In clinical practice this is usually determined by measuring the Cobb angle in coronal views of X-ray scans~\cite{Aebi05},  however measuring scoliosis in the supine position has also been shown to be possible \cite{taylor_identifying_2013,Jamaludin19a} and MRI can be useful for understanding disease etiology and symptoms \cite{ozturk_role_2010}. This is a more difficult task in sagittal scans as it requires sensitive detection of the sides of vertebra with a clear decision boundary as to when a vertebra is present or not present in a slice which can be difficult in cases of partial visibility. In the entirety of the Genodisc dataset, scoliosis was reported by a radiologist in 198 of 3542 scans, across 2009 patients. By measuring statistics of a quintic polynomial fit through vertebra centroids and using them as predictive features we develop classifiers for this label. Specifically, we measure the maximum curvature of the polynomial, and the maximum deviation of the curve from a straight vertical line fit through the vertebrae, assuming that a low curvature vertebral column with little deviation from the centreline corresponds to a non-scoliotic spine. 
While ground truth scoliosis labels were not available for the whole spine scans, we give qualitative results, comparing the features of curves fit though vertebrae of a scoliotic and non-scoliotic scan from Zuki\'c. Results of these experiments are shown in Figure~\ref{fig:scoliosis}.

    \noindent\textbf{Discussion: }The results for automated scoliosis detection from sagittal scans are promising. Using simple classifiers AUCs of 0.636-0.689 are achieved in a highly class-imbalanced problem. Of the features measured, distance from the vertical centreline of the vertebrae performed best. The system is also shown to  capture scoliotic curves in full spine scans too. These experiments illustrate that the proposed method produces a strong geometric representation of the spine, which can be used in further downstream tasks.
	\section{Conclusion}
 We introduce a novel method for vertebrae detection and labelling in
 whole spine sagittal MRIs. It shows state-of-the-art results for vertebra identification in 
 lumbar scans,  with little performance drop in the whole spine
 case, 
and is robust to a range of spinal defects, numerical
variations of vertebrae and different scanning protocols.  
We also
 demonstrate a potential diagnostic application: automated detection
 of scoliosis from sagittal MRIs. 
Future work will include
 integrating automated detection of different spinal pathologies and
 implementing the system for CT scans.
\smallskip	 

\noindent\textbf{Acknowledgements.}
The authors would like to thank Dr.\ Sarim Ather for useful discussions on spinal anatomy and
clinical approaches to diagnosing disease, as well as assistance labelling the data.
Rhydian Windsor is supported
by Cancer Research UK as part of the EPSRC CDT in Autonomous Intelligent Machines and Systems (EP/L015897/1).
Amir Jamaludin is supported by EPSRC Programme Grant
Seebibyte (EP/M013774/1). The Genodisc data was obtained during the
EC FP7 project GENODISC (HEALTH-F2-2008-201626).

\bibliographystyle{splncs04}
\bibliography{shortstrings,vgg_local,new_references,vgg_other}
\end{document}


\begin{center}

\section*{Appendix}
\end{center}

\subsection*{1. Overview of Proposed Pipeline}
\vspace{-5mm}
\begin{figure}[h!]
    \centering
    \includegraphics[width=0.80\textwidth]{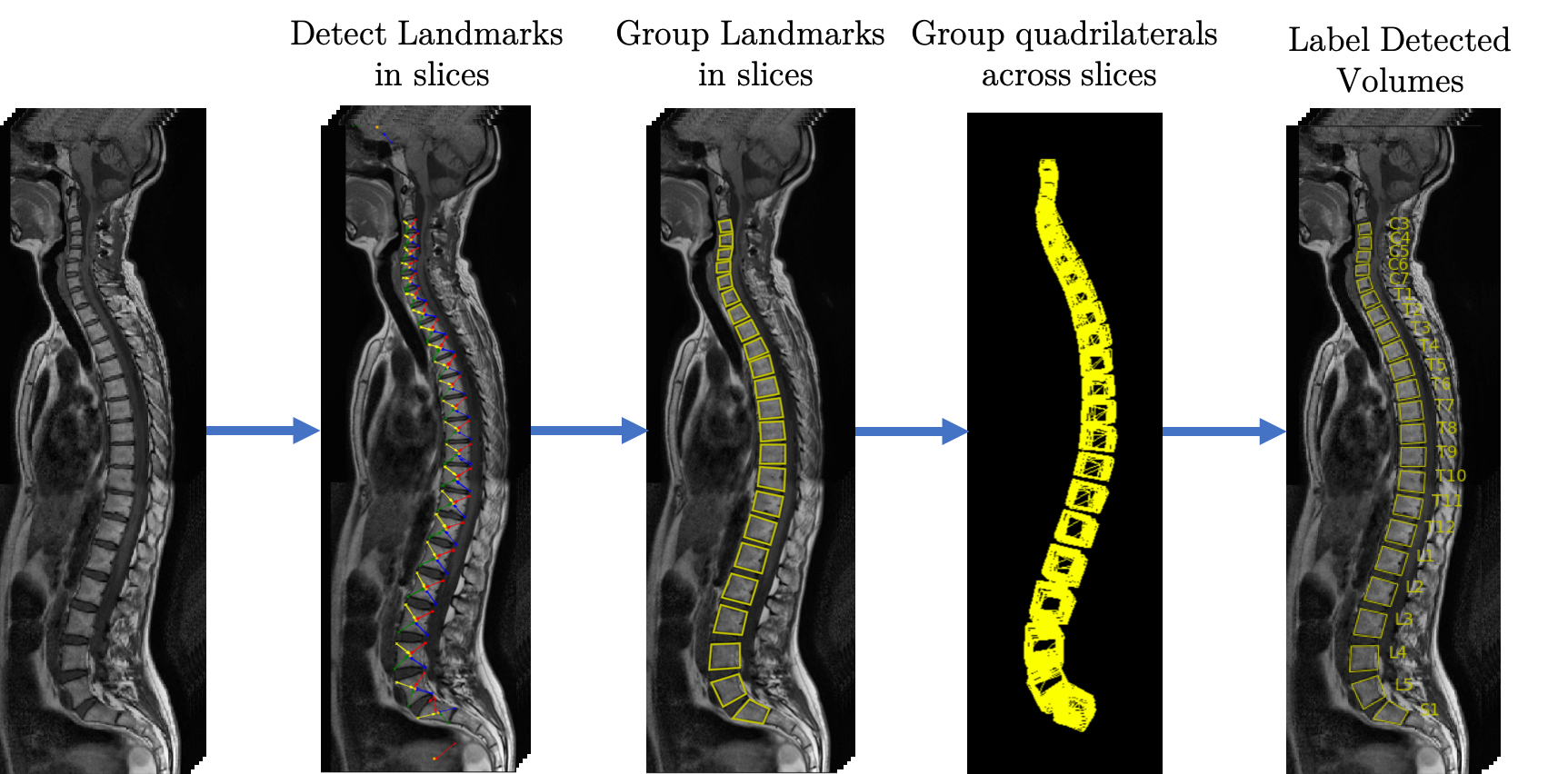}
    \caption{An overview of the proposed detection and labelling pipeline. Vertebrae landmarks are detected, grouped into quadrilaterals in each sagittal slice and finally grouped across slices. The output volumes are input into the labelling pipeline.}
    \label{fig:my_label}
\end{figure}

\subsection*{2. LSTM Baseline Architecture}

\begin{figure}
    \centering
    \includegraphics[width=.7\textwidth]{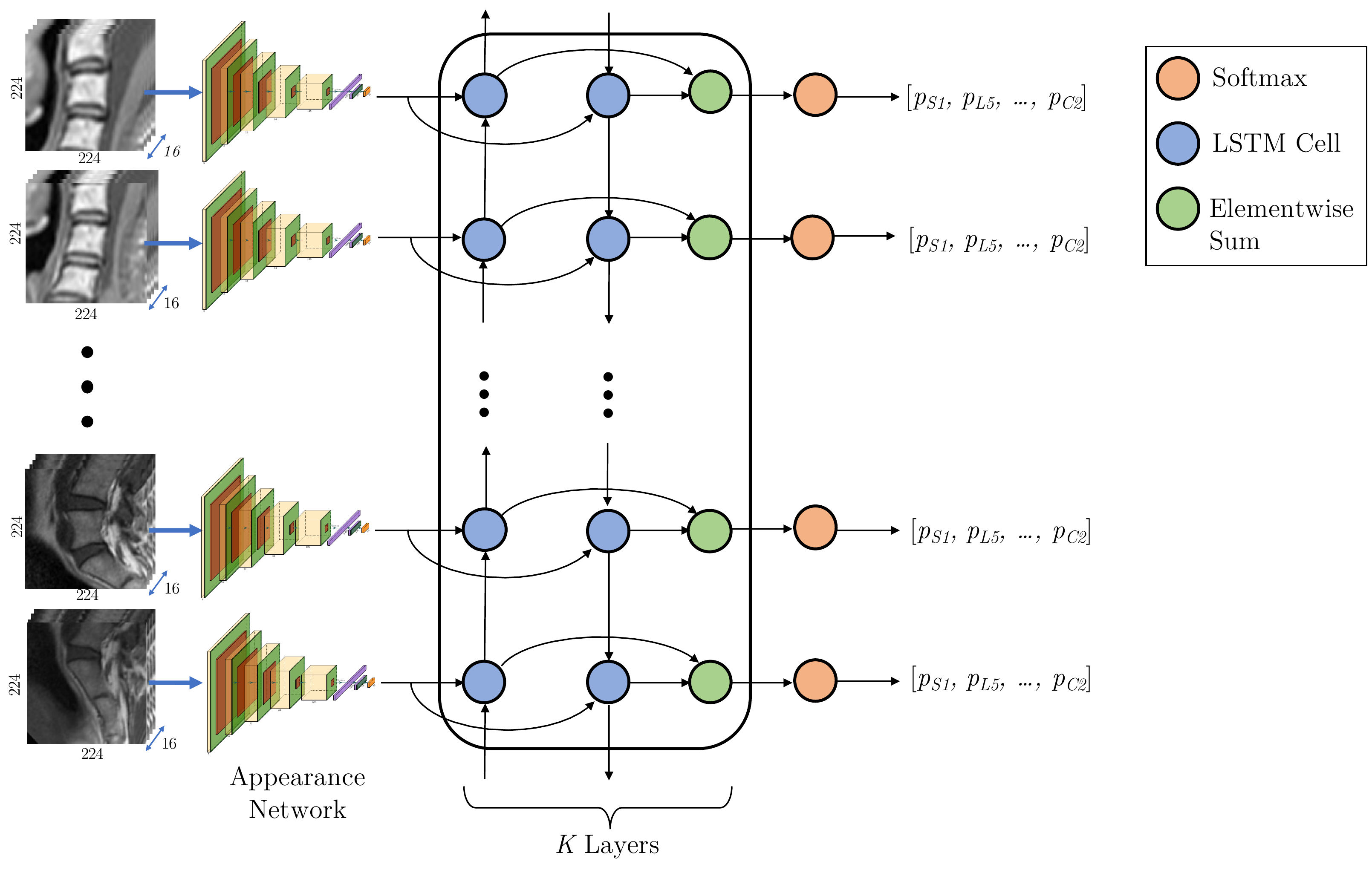}
    \caption{The LSTM architecture used as a labelling baseline. Output probability maps from the appearance network are used as input to a bidirectional LSTM, with the inputs ordered by the height of the vertebrae in the image. The architecture outputs assigns a probability vector for each detection which the likeliness of the vertebra being each level from S1 to C2. In the experiments in Section 4, $K=2$ is used.}
    \label{fig:my_label}
\end{figure}

\FloatBarrier
\newpage

\subsection*{3. Other Reported Results}
\vspace{-.5mm}
	\begin{table}[!htp]
		\centering
		\caption{Reported results of other methods proposed for the segmentation of vertebrae in CT and MR spinal scans evaluated on different datasets. We attempted to access a publically available dataset from [2] although did not get a reply prior to submission.}
	    {\fontsize{9}{11}\selectfont
		\begin{tabular}{|c|p{2.8cm}|p{5.3cm}|c|c|r@{}l|c|}
			\hline
			Method & FoV &  Comment & Prec. (\%) & Rec. (\%) & \multicolumn{2}{c}{LE(mm)} & IDR (\%) \\
			\hline
			Cai et al.~[2] & Arbitrary &  MR \& CT, Deformable shape model& - & - & 2.54-&3.12 & 92-97 \\
			Forsberg et al.~[4] & Lumbar, Cervical & MR, requires S1 or C2 visibility & 99.6-100 & 99.1-99.8& 1.18-&2.60&96-97\\
			Glocker et al.~[4] & Mostly Abdominal & CT, Random Forest Classifier & - & - & 5.94-&8.53& 72-85 \\
			Glocker et al.[5] & Mostly Abdominal & CT, Random Forest Classifier & - & - &7-&14.3 & 62-86\\
			Yang et al.~[18]& Arbitrary & MR, LSTM labelling & - & -& 6.9-&9.0 & 83-89\\

			\hline
		\end{tabular}
		}
		\label{tab:comparison_table}
		
	\end{table}
\FloatBarrier
\subsection*{4. Detection, Appearance and Context Network Architecture Diagrams}

\begin{figure}[!ht]
\centering
\subfloat[The detection network. All layers have $3\times3$ kernels with $1\times1$ padding. MaxPool operations have a kernel size of $2\times2$. \label{fig:subim2}]{%
  \includegraphics[width=.3\textwidth]{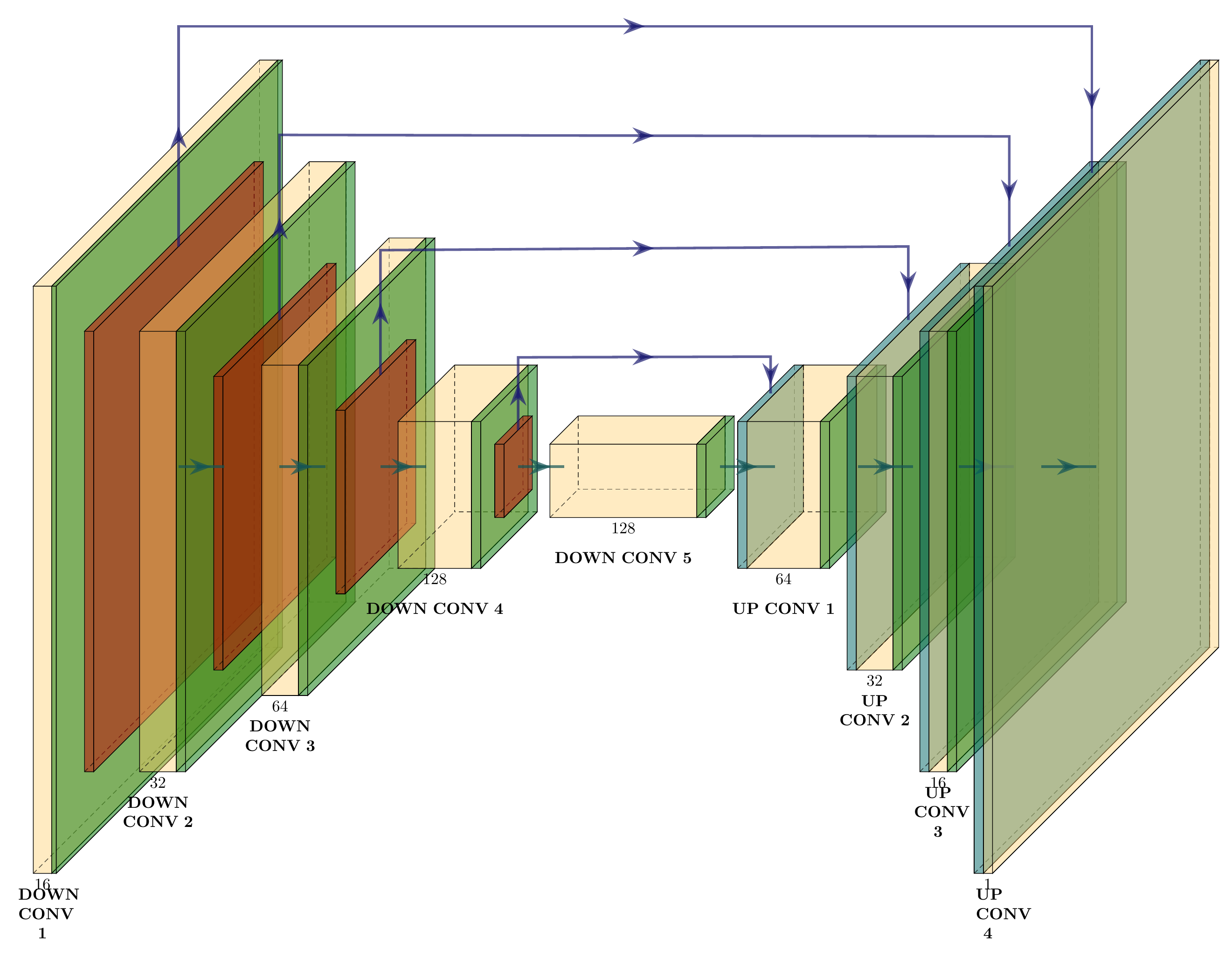}%
}\hfil
\subfloat[The appearance network used to label vertebrae volumes. The first two layers have kernel sizes of $3\times3\times1$ with $3\times3\times3$ in later layers.\label{fig:subim1}]{%
  \includegraphics[width=0.3\textwidth]{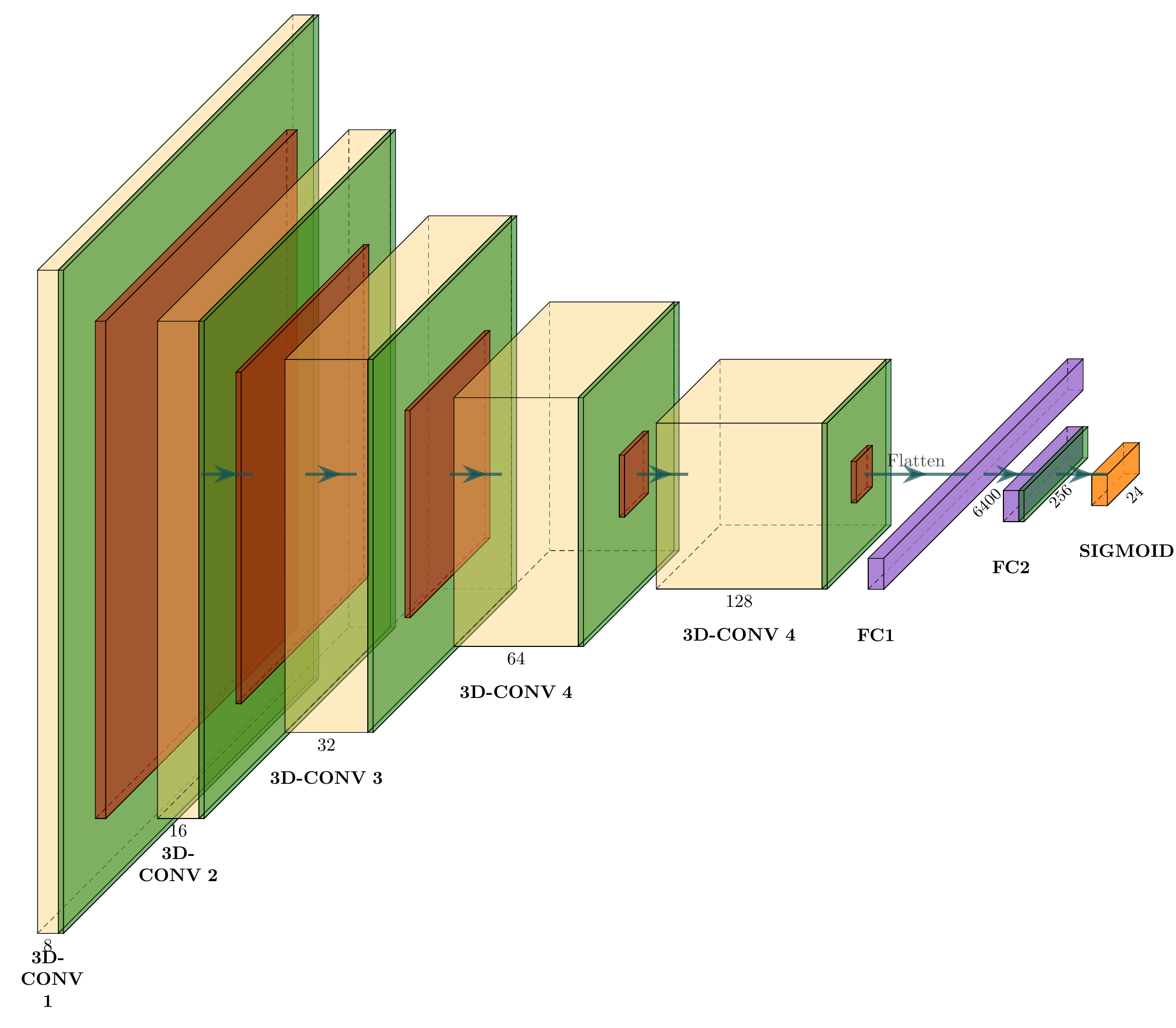}%
}\hfil
\subfloat[The context network. All Conv layers have $5\times3$ kernels with $2\times1$ dilation, $4\times1$ padding and $2\times1$ stride. MaxPool operations have $2\times1$ kernels.\label{fig:subim2}]{%
  \includegraphics[width=0.3\textwidth]{images/ContextNetwork.pdf}%
}

\caption{The network architectures used in the detection and labelling pipeline.Yellow blocks indicate convolutional layers, green blocks indicate batch normalisation, purple layers are fully connected and orange layers are sigmoid activation units. All layers except the last use ReLU activation units (not shown).}
\label{fig:image2}

\end{figure}
\FloatBarrier
\newpage
\subsection*{5. Detection and Labelling Accuracy Across Different Vertebral Levels}
\begin{figure}[!ht]
    \centering
    \includegraphics[scale=0.7]{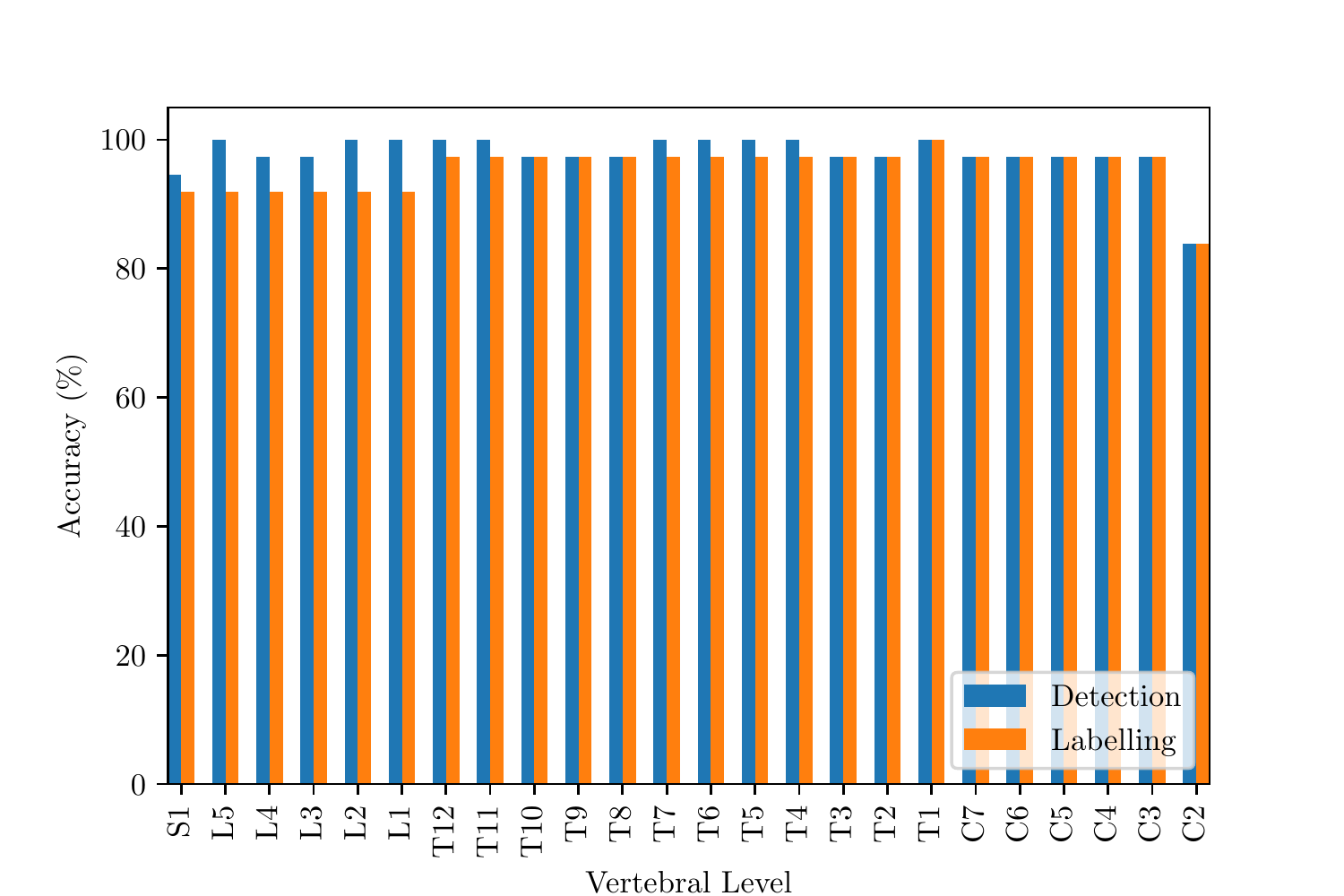}
    \caption{Detection recall and vertebral identification rate across different vertebral levels in the OWS whole spine dataset. The system is least accurate at detecting vertebrae at the extremities of the vertebral column (S1 and C2).}
    \label{fig:my_label}
\end{figure}